# Ultra-narrow linewidth self-adaptive photonic oscillator: principle and realization

Mehdi Alouini, Gwennael Danion, and Marc Vallet

*Univ. Rennes, CNRS, Institut FOTON - UMR 6082, F-35000 Rennes, France*

mehdi.alouini@univ-rennes1.fr

**Abstract**

Highly coherent optical sources are a key element in several fields of physics, in particular in time frequency metrology. Over the past decennia, there has been particular efforts in developing such sources to the expense of sophisticated laser systems and relatively smart electronics. We propose here a new general principle of a self-adaptive oscillator where the intricate operation of a 100-m-long active optical resonator and a standard semiconductor laser offers a very high spectral purity and can be tailored to any wavelength. Single frequency operation of this self-adaptive photonic oscillator is achieved without any servo locking or stabilization electronics. Free running operation leads to a Lorentzian linewidth of 40 mHz. The long-term drift of the optical frequency in the free running regime is within 10 MHz over hours. This principle applies to any wavelength attainable by laser diodes which opens tremendous opportunities in particular in applications where atomic or molecular transitions require precise wavelengths.

## 1. Introduction

Owing to their major role in most modern physics experiments, single-frequency narrow-linewidth lasers are the subject of intensive research efforts. They are becoming mandatory for a wide field of applications including microwave photonics[1,2,3], time-frequency metrology[4], atom manipulation[5], atomic clocks[6,7,8,9], high resolution spectral analysis, resonant gyroscopes, coherent communications[10,11,12], Lidar[13], coherent long-haul sensing, gravitational wave detection[14,15,16], quantum optics, and more generally in research involving light-matter interactions[17]. Over the past decades, several approaches have been pursued in view of generating the most coherent possible electromagnetic emission. In stringent applications such as atom manipulation or time-frequency metrology, external cavity semiconductor lasers are currently widely used due their scalability in terms of wavelength, the external cavity configuration improving their frequency noise performances[18,17]. Such lasers are then often stabilized on Fabry-Perot cavities[19,20,21], atomic transitions[22] or optical frequency combs[23,24]. As compared to monolithic semiconductor lasers, solid-state[25,26,27] and fiber[28,29]. lasers offer intrinsic narrow linewidths. Moreover, these lasers are free from amplitude to phase noise conversion, which make them appreciated in demanding applications where stable and high spectral purity is necessary such as for gravitational wave detection[14,15]. Among solid state lasers, Brillouin fiber lasers have attracted a peculiar attention[30,31,32,33,34]. Indeed, the Stimulated Brillouin Scattering (SBS) offers extremely narrow gain bandwidth which facilitates single mode oscillation of a long optical resonator favoring a natural narrow linewidth[35].

In addition to those inherent narrow linewidth lasers, active or passive stabilization schemes are usually employed to further stabilize the laser line and/or to narrow down its linewidth. In this regard, self-referenced frequency combs became the standard and undoubtedly the most powerful stabilization tool that is employed as a ruler which enables to stabilize the single mode laser frequency and to narrow down its linewidth at the same time. Obviously, sophisticated phase locked loops whose bandwidth is higher than the laser linewidth must be employed. Alternatively, locking the single mode laser to an UltraLow Expansion glass cavity (ULE) that is acoustically insulated, thermalized, and put into vacuum has been shown to be a very attractive approach as well[20,21]. Indeed, as compared to atomic or molecular lines the ULE cavity offers thousands of possible stable optical lines. A Pound-Drever-Hall servo loop is then commonly used to lock the laser frequency to one of the ULE cavity resonances. In all cases single mode laser has to exhibit a narrow linewidth in order to release the stringent constraints on the servo-locking electronics. Accordingly, the development of compact and ultra-narrow linewidth lasers is still today a very hot research topic[36]. With this aim in view, several linewidth narrowing techniques have been demonstrated more recently. They generally rely on optical reinjection of the single mode laser using external Bragg grating reflectors, micro-resonators[37] or even Rayleigh scattering in long fibers[38,39,40,41].

In parallel, in the microwave domain, an important step has been taken in achieving high spectral purity microwave oscillators[42]. Accordingly, Opto-Electronic Oscillators (OEO) are to date the simplest, most efficient and reliable microwave resonators. They rely on the use of long-fiber optical delay lines in order to bring the microwave resonator length to kilometer scale[43]. Single mode operation is ensured by a microwave filter, whereas the linewidth is ruled by the long fibered delay line[44,45]. Interestingly, OEOs do not require any servo locking electronics as the optical delay line is part of the oscillator itself.

We report here on a Self-Adaptive Photonic Oscillator (SAPO)[46] whose operation principle in the optical domain is inspired by the OEO paradigm in the microwave domain. Accordingly, the extremely high spectral purity obtained in the free running regime, without any servo-locking, contrasts with any other reported approach for generating narrow linewidth optical waves.

## 2. SAPO principle

Extending the OEO principle in the optical domain relies on the possible design of a long resonant loop which operates itself in the optical domain. Fig. 1 illustrates the principle of an OEO in comparison to what should be a SAPO. The main ingredients for realizing an OEO are (i) a filtering stage which selects the central microwave frequency (ii) a long optical delay line which provides the high spectral purity when embedded within the microwave resonator, and (iii) the amplifying stage which compensates for the losses mainly generated by microwave-to-optical and optical-to-microwave conversions. It is worthwhile to notice that a narrow bandwidth microwave filter is mandatory in order to ensure single frequency oscillation. By moving to the optical domain, i.e. at around 200 THz frequencies, this constraint becomes a challenging aspect.

SBS effect is the cornerstone for building a SAPO. Indeed, this physical effect, when induced in a single mode silica fiber, enables optical amplification over an extremely narrow bandwidth, i.e. of around 30 MHz[47,48]. Accordingly, the filtering function as well as the amplification function can be achieved together. Nevertheless, SBS requires, firstly, a sufficiently long time/distance light-matter interaction and needs, secondly, to be activated with a high intensity and narrow linewidth pump beam. To this aim, Brillouin oscillation is usually realized in micro-resonators in order to take advantage of their high-quality factor or in long fiber resonators provided that their free spectral range is lower than the inherent 30 MHz SBS gain bandwidth. In all cases the high spectral quality of the pump beam is a key condition. More importantly, an electronic servo-locking apparatus is required to efficiently feed the Brillouin resonator[32,49].

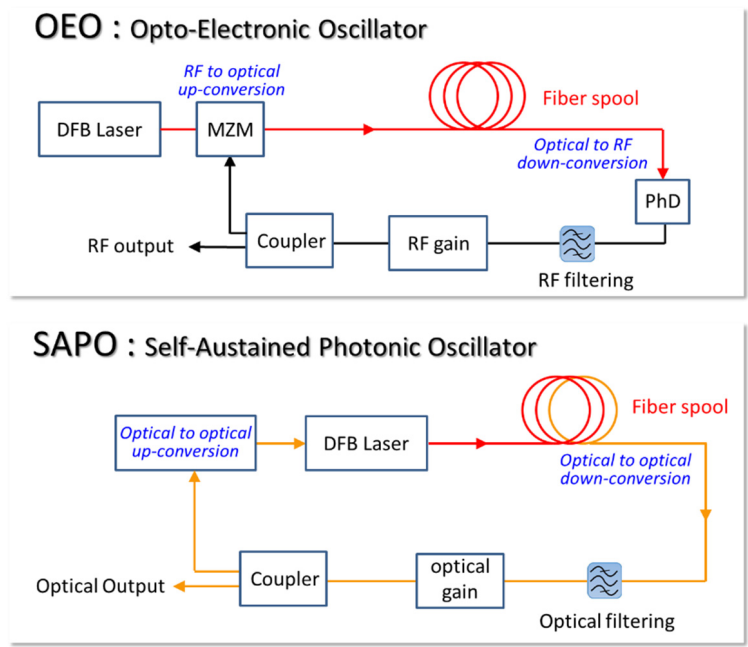

**Fig. 1:** Parallel between the operation principle of an OptoElectronic Oscillator (OEO) and a Self-Adaptive Photonic Oscillator (SAPO). The SAPO can be considered as an OEO extension in the optical domain.

For realizing an SAPO, these two constraints must be released. Let us consider that the pump laser and the Brillouin resonator constitute respectively the primary and secondary optical resonators, where the primary resonator provides the required energy, and the secondary long resonator generates a single mode high spectral purity electromagnetic field. To make the SAPO operation possible the two resonators must be optically intertwined so that they behave from an electromagnetic point of view as a single resonator. This *single-resonator* condition implies, firstly, that the secondary long resonator must be open (i.e. non-resonant) regarding the field coming from the primary resonator. Conversely, it has to be closed (i.e. resonant) for the generated electromagnetic field so as to provide the targeted high spectral purity. This can be fulfilled using a non-reciprocal resonator which is, by construction, non-resonant clockwise regarding the incoming field but resonant counterclockwise. The *single-resonator* condition implies also that the electromagnetic field generated by the second resonator must be able to easily couples into the primary resonator. Thus, counterintuitively, the primary resonator has to exhibit a very low-quality factor. This property is commonly satisfied in semiconductor lasers. Consequently, the overall constraints to realize the proposed SAPO have practical solutions, making it conceivable form a technical point of view.

## 3. SAPO practical realization

In practice the above-mentioned ingredients to realize a self-adaptive photonic-oscillator can be gathered together starting from a standard single mode DFB semiconductor laser which pumps a long Brillouin resonator designed to be non-resonant for the pump field but resonant for the Stokes wave as depicted in Fig. 2(a). The generated contra-propagating Stokes-wave, which in turn experiences the resonant condition, presents a high spectral purity. This Stokes wave is then reinjected to the DFB laser after being frequency up-shifted to match the pump central frequency (Fig. 2(b)). Provided that this frequency shift is precisely equal to the Brillouin offset, the DFB laser frequency will acquire the spectral properties of the reinjected field and the resonant condition of the whole photonic oscillator is accordingly satisfied. Hence, after one round-trip the linewidth of the DFB laser narrows down making

July 7, 2021Ultra-narrow linewidth self-adaptive photonic oscillator: principle and realization   M. Alouini et al.   3

the Brillouin process even more efficient leading to a narrower Stokes-wave linewidth which is going to reinject the DFB laser once again, and so on.

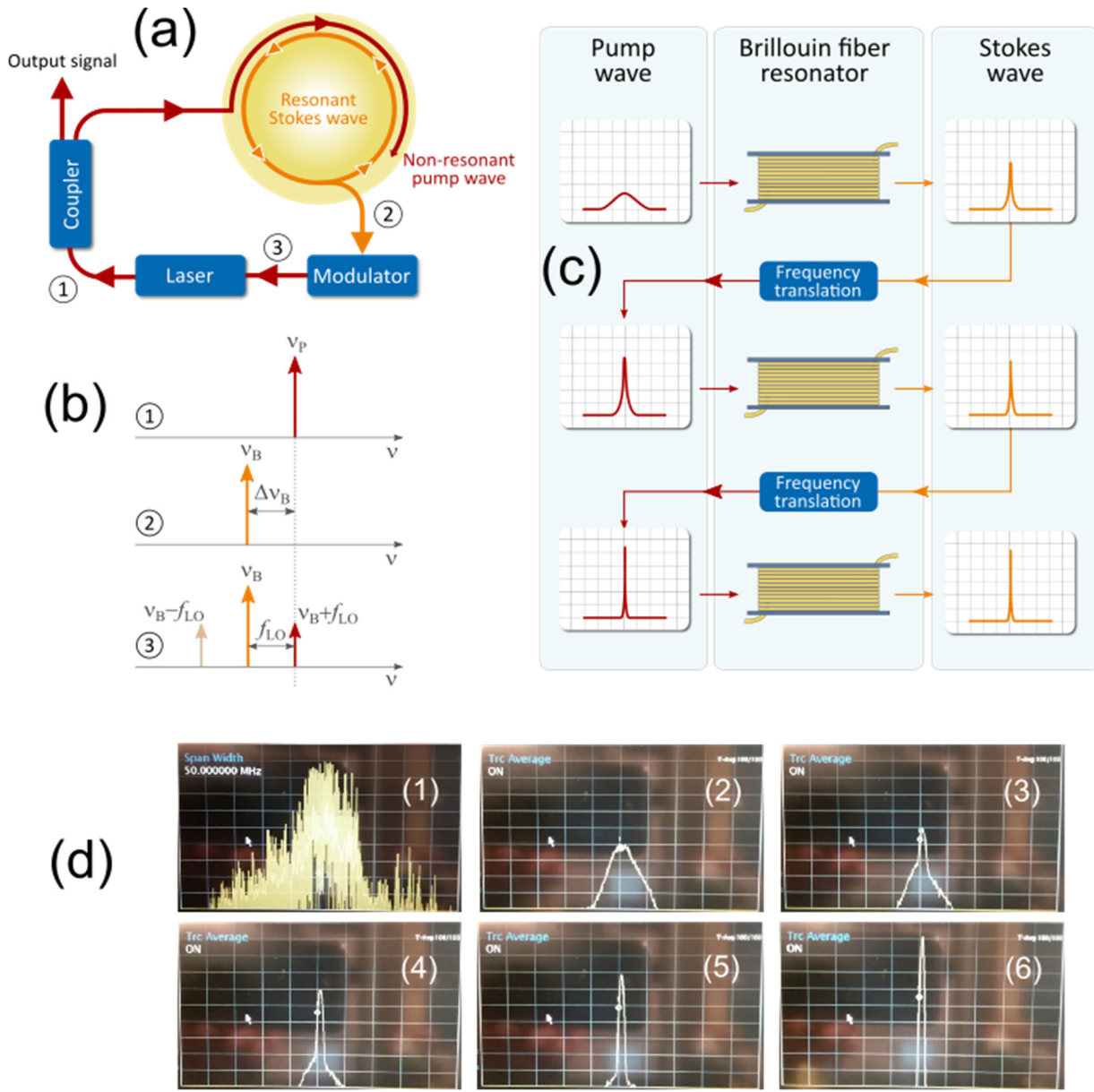

**Fig. 2: SAPO operation principle.** **(a)** SAPO architecture. The modulator is modulated at the mean Brillouin phonon frequency $f_{LO}$. **(b)** The non-resonant pump, at $\nu_P$, generates a resonant Stokes wave, at $\nu_B$, which is then up-converted to $\nu_B+f_{LO}$. $\nu_B+f_{LO}$ being nearby $\nu_P$, the pump locks to this frequency. **(c)** Cascade effect leading to the linewidth narrowing process. As the pump is optically reinjected by $\nu_B+f_{LO}$, its linewidth narrows down enhancing the Stokes wave generation process. After a few round trips the pump and the resonant Stoke waves get the same spectral purity. **(d)** Real time screen capture of the beatnote between the pump and the Stokes wave, $\nu_P-\nu_B$, observed with an electrical spectrum analyzer. (1) and (2), the up-converted Stokes wave is not reinjected into the DFB pump laser. (2) 100 times average of (1). From (3) to (6), time evolution of the beatnote when the up-converted Stokes wave is reinjected into the DFB pump (with 100 times average). The beatnote shrinks while its power spectal density increases showing that the energy originally contained in (2) concentrates in this narrow line.

This cascade effect, schematized in Fig. 2(c), takes place inherently. It implies a dramatic linewidth-reduction of the DFB laser while the Brillouin process efficiency is enhanced. After a few round-trips, the SAPO reaches its stationary regime where the DFB and Stokes-wave linewidths are equal.

In practice, the 100-m-long Brillouin loop is closed using an optical circulator that ensures a resonant condition for the counter-clock-wise Stokes-wave while offering a single pass condition for the clockwise pump wave. This enables the whole pump energy to enter into play whatever the initial pump linewidth. In our proof of concept, the DFB laser oscillates at 1550 nm and delivers 20 mW output power. Half of its power is sent into a home-made low power erbium doped fiber amplifier which raises the output power to 100 mW before entering the fiber loop. The remaining power 10 mW serves as useful signal. The free-running linewidth of this laser has been measured to be 4 MHz, i.e., four times larger than the Free Spectral Range (FSR) of the nonreciprocal Brillouin resonator. Owing to the homogeneous nature of SBS gain, single frequency and mode hop free operation of the Stokes wave can be obtained provided that one resonance of the Brillouin resonator matches the maximum of the SBS gain[50]. This condition is fulfilled by up-shifting the Stokes-wave frequency by the amount corresponding to the Brillouin frequency offset, i.e., 10.9683 GHz in our case. To this aim, part of the Stokes wave is extracted and phase modulated at 10.9683 GHz before being reinjected to the DFB laser (Fig. 2(b)). Among the two generated sidebands, the upper sideband frequency will coincide with the pump frequency. The two remaining tones being far from the DFB laser oscillating frequency, they are filtered out. Thus, the DFB laser locks to the up-shifted Stokes-wave and acquires its spectral purity (Fig. 2(d)). More importantly, this DFB-laser being itself the pump source generating the Stokes-wave, the locking mechanism is self-adaptive as it will be discussed later. The oscillation process evolves after a few round-trips towards a stationary state where the pump carries the same spectral purity as the Stokes wave without any servo-locking. When turned on, the SAPO do not exhibit any mode hopping or frequency instabilities or intensity fluctuations as long as it is turned on, i.e., during days.

## 4. Free running spectral characteristics

**Spectral purity**

The SAPO has been assembled in a compact and thermalized housing (Fig. 3(a)). The frequency noise is measured using a delayed self-heterodyne apparatus based on an unbalanced Mach-Zehnder interferometer including a 80 MHz acousto-optic modulator on one and a 700 m long fiber on the other arm. The self-heterodyne beatnote spectrum is then measured around 80 MHz with an electrical spectrum analyzer. An overview of the self-adaptive linewidth narrowing is given in Fig. 2(b) where the self-heterodyne beatnotes, for the DFB outside and inside the SSPO, are displayed with the same scales for a span of 150 MHz. The latest beatnote is zoomed in in Fig. 2(c) up to a span width of 50 kHz.

In order to extract comprehensive results this beatnote is now measured with a Rohde & Schwarz RS-FSW26 phase noise analyzer. The acquired phase noise is then converted into frequency noise taking into account the interferometer response[51,52]. Moreover, to cover the 0.01-100 kHz bandwidth, and due to the low levels of noise obtained here, it was necessary to acquire the phase noise using two fiber lengths in the unbalanced interferometer, that is 700 m and 5 km for respectively the high and low frequency parts of the spectrum. As shown in the composite spectrum of Fig. 3(d), the SAPO frequency noise is 100 Hz$^2$/Hz, 2 Hz$^2$/Hz and 0.03 Hz$^2$/Hz at 0.1 kHz, 1 kHz and 10 kHz frequency offset respectively. This spectrum indicates that the SAPO intrinsic linewidth is at around 40 mHz. Moreover, the β-separation line, conventionally used to assess the flicker noise[53], reveals a 200 Hz linewidth only for 0.1 s integration time. This result is the state of the art for a free-running and servo-free oscillator. It confirms the high potential of the proposed SAPO principal for realizing ultranarrow linewidth optical sources.

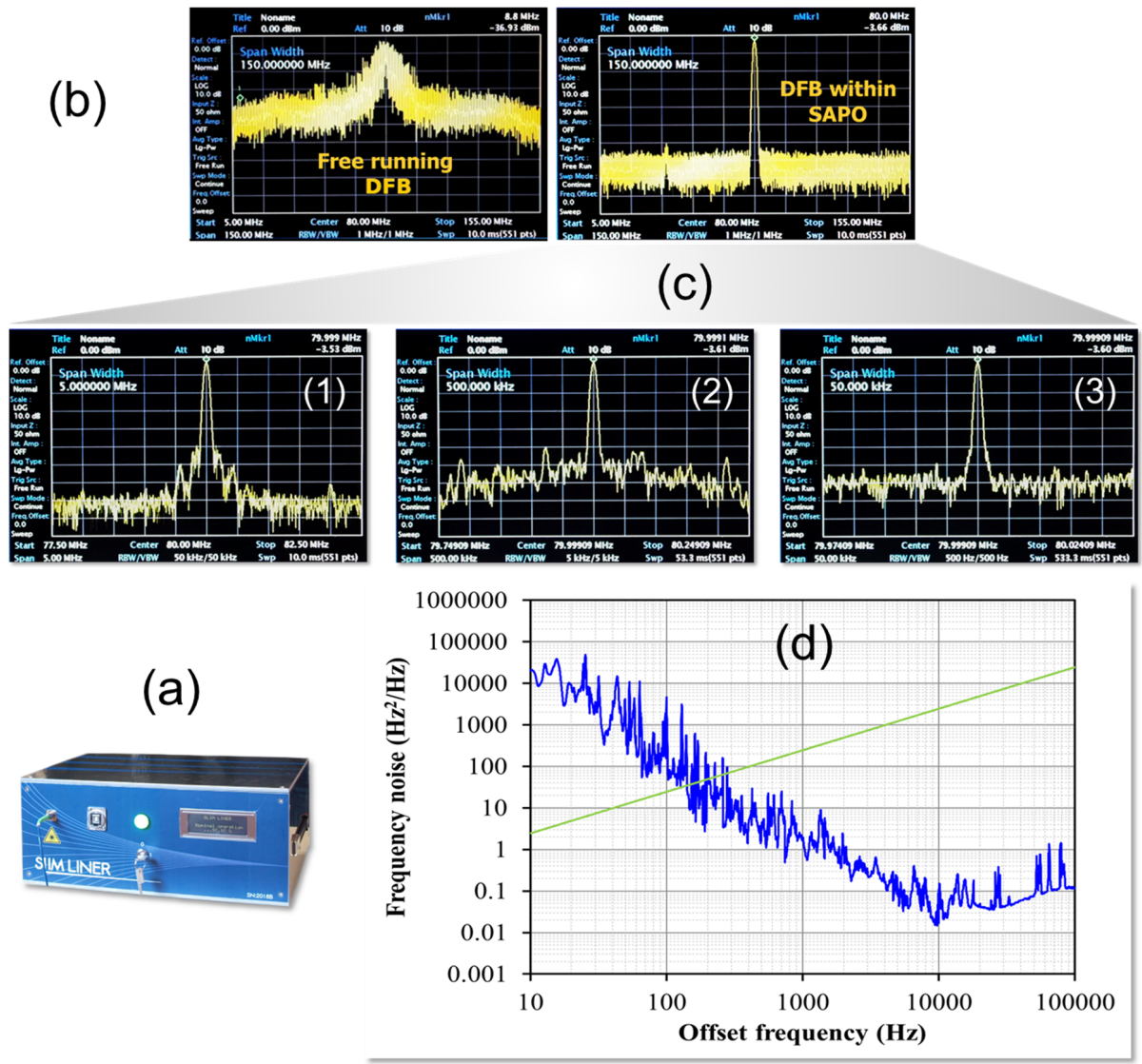

**Fig. 3: Spectral purity.** **(a)** The SAPO has been assembled in a compact thermalized housing prior to noise characterization. **(b)** Self-heterodyne spectra of the DFB laser outside the SAPO (left-hand side) and when it is imbedded in the SAPO (right-hand side). **(c)** Same as (b) with decreased spectral analysis span width: (1) 5 MHz, (2) 500 kHz and (3) 50 kHz. **(d)** Frequency noise computed from the phase noise of the self-heterodyne signal. The straight green line accounts for the β-separation line which reveals a flicker noise linewidth of 200 Hz for 0.1 s integration time. The natural linewidth deduced from the frequency noise floor is 40 mHz.

**Intensity noise characteristics**

The SAPO includes by principle an extremely long delay line imbedded into the resonant system. Taking into account the propagation length in the SAPO, including the Brillouin loop, the effective photon lifetime is predicted to be in the μs range. This photon lifetime being significantly larger than any other time constant of the system, class A operation is expected[54], meaning that the SAPO intensity noise should not exhibit any resonant excess noise and should be shot noise limited. The relative intensity noise (RIN) has been measured at high (Fig. 4(a)), medium (Fig. 4(b)) and low (Fig. 4(c)) frequencies using dedicated RIN benches[55].

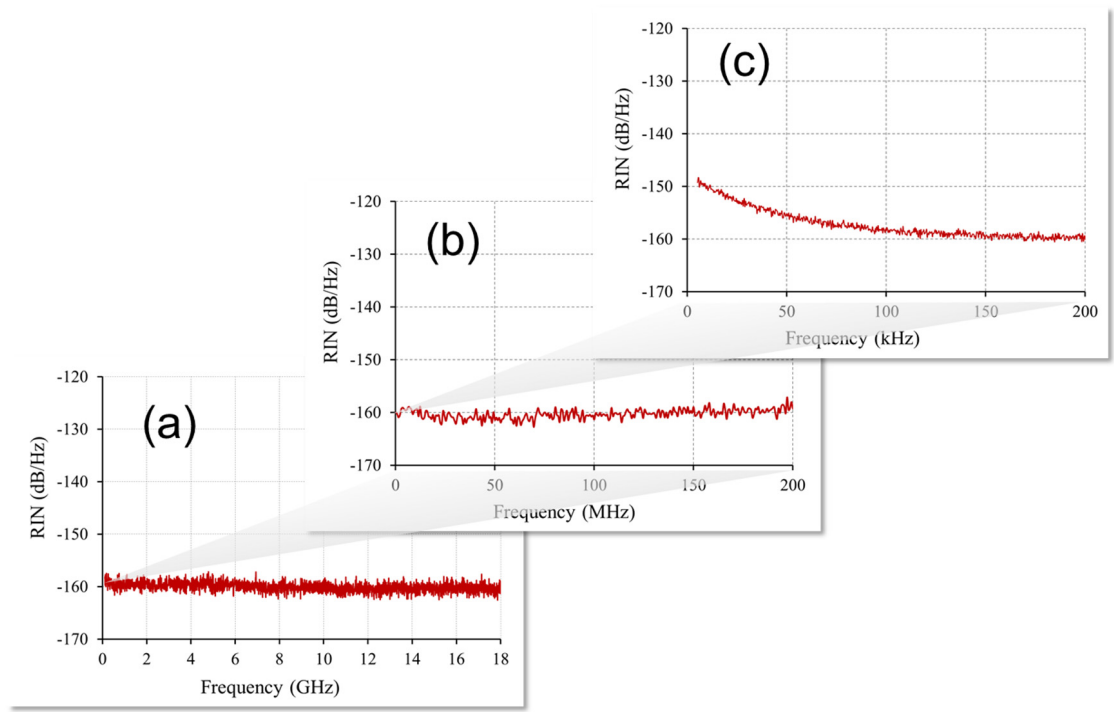

**Fig. 4: Intensity noise.** Relative intensity noise (RIN) spectra of the SAPO at (a) high frequency, (b) medium frequency and (c) low frequency. The SAPO intensity noise sticks to the shot noise level of -160 dB/Hz for 5 mW photo-detected power. Below 100 kHz, the optical noise is hidden by the electronic noise of the detection chain.

As shown in the different RIN spectra of Fig. 4, the SAPO exhibits a white noise whose level is -160 dB/Hz from 100 kHz to 18 GHz. This level corresponds to the shot noise floor for 5 mW detected optical power. It is worthwhile mentioning that we have deliberately limited the detected power to 5 mW in order to ensure that the RIN measurement chain operate within its linear region. Even though shot noise limited operation was not the main goal for developing this SAPO, it actually overcomes any other class A laser reported in the literature[56]. Indeed, an equivalent optically pumped class A laser would show an excess noise plateau due to the pump noise up to 100 kHz[57].

**Frequency stability**

In order to characterize the SAPO long-term frequency drift in the free running regime, a second identical SAPO has been built. Their Brillouin-fiber-loop is thermalized using a thermoelectric controller. Moreover, they are properly insulated from acoustic vibrations. The two SAPOs have been positioned in two different experiment rooms so that they experience uncorrelated environmental perturbations. By adjusting the temperature of fiber spool resonator, the wavelength of the second SAPO is set at 0.16 nm, i.e. 20.2 GHz, from the first SAPO. Their beatnote is then acquired using a fast photodiode connected to the electrical spectrum analyzer. A beatnote spectrum between the two SAPOs with a span width of 300 kHz is illustrated in Fig. 5(a). Owing to the fact that the two SAPOs are free running, a random walk ranging from 10 to 20 MHz is observed for an acquisition time of 4 hours. The standard Allan deviation is also reported in Fig. 5(b). It can be noticed that it reaches a plateau above an integration time of 100 s. To go further in terms of characterization, the coming step will be the stabilization of such SAPO on molecular transitions.

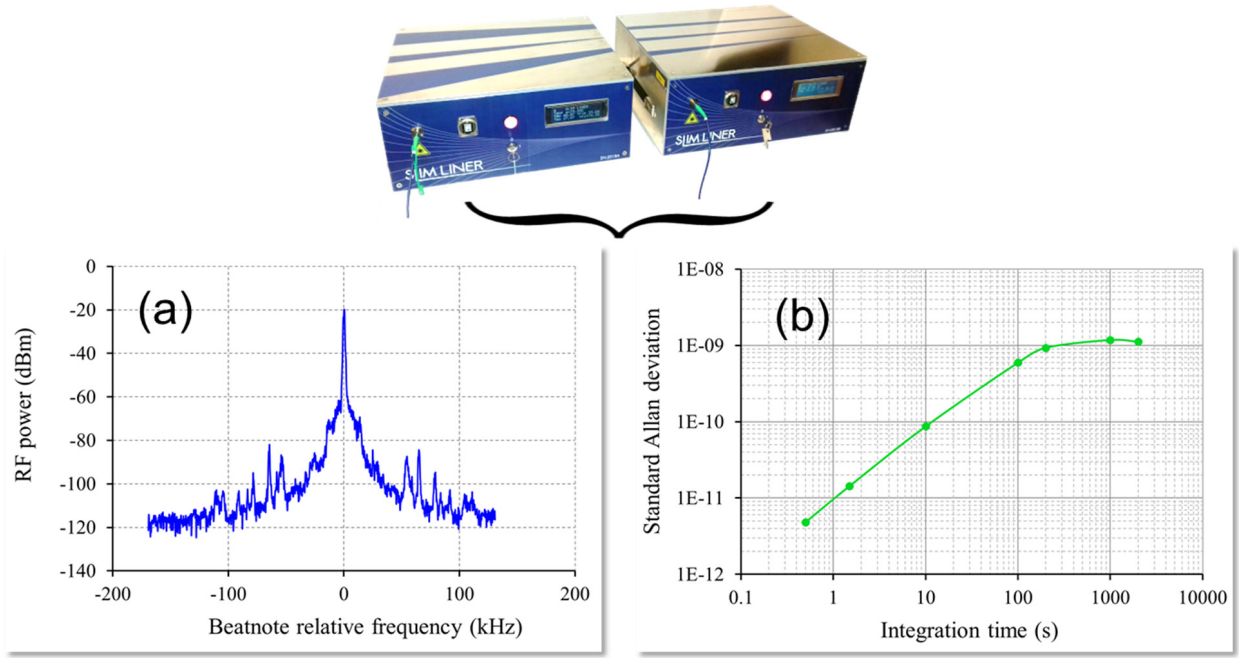

**Fig. 5: Frequency stability.** Two identical SAPOs have been assembled in order to assess the long-term frequency stability in the free running regime. **(a)** Beatnote between the two independent SAPOs observed with an electrical spectrum analyzer (span width of 300 kHz). **(b)** Standard Allan deviation computed from the frequency drift of the beatnote (a).

## 5. Discussion on SAPO distinctive characteristics

**Why Stimulated Brillouin Scattering?**

The first idea that comes in mind to obtain a narrow linewidth photonic oscillator is to realize the longest possible laser cavity. Indeed, the Schawlow Townes linewidth, i.e. the quantum limited linewidth, is inversely proportional to the photon lifetime inside the cavity. The longer the cavity, the narrower the linewidth is. Unfortunately, when the laser cavity exceeds a few meters, single mode operation becomes difficult to ensure due to the lack of sufficiently narrow and stable filters that can be inserted into the laser cavity. Furthermore, assuming that such a filter exists, an active tracking would be necessary to avoid mode hopping. This major constraint explains the fact that current techniques for narrowing down laser linewidths involve external delayed filtering components that are not part of the optical oscillator itself. With this respect, embedding the SBS effect within the optical oscillator brings the required narrow linewidth optical filtering function and also the self-adaptive function releasing the constraint of servo-locking. Finally, owing to the homogeneous spectral broadening of the SBS gain[50], the length of the Brillouin resonator can exceed 10 times that which provides one mode within the Brillouin gain. As a result, a Schawlow Townes linewidth below 1 µHz is reachable.

**Why a large linewidth a pump laser?**

In the proposed SAPO architecture the pump laser must be efficiently sensitive to the Stokes wave, that is fed back, in order to act as part of the overall oscillator. Consequently, its intrinsic linewidth must be wide enough to make the SAPO oscillation robust against environmental perturbations. This condition is fulfilled by any standard semiconductor lasers. In return the Brillouin resonator has to be non-resonant for the pump so that any change in the spectral characteristics of the pump do not affect the pumping efficiency, bringing the SAPO to a self-adaptive oscillation.

**Why SAPO operation contrasts with standard frequency-locked lasers?**

The frequency robustness of the SAPO relies on the interleaved interaction between the DFB laser and the active Brillouin resonator. Its behavior cannot be understood considering these two elements separately. Indeed, let us assume that the frequency of the Stokes wave drifts due to a change of the fiber optical index. The DFB pump laser being injected by the Stokes wave, its frequency, and consequently the maximum of the induced SBS gain, will automatically follow this drift (Fig. 6(a)). As a result, no mode hops occur. Such a frequency dragging effect can be seen in Fig. 6(b) where the temperature of the Brillouin fiber was linearly increased. The SAPO frequency changes continuously over a span of 1 GHz, that is 500 times the fiber resonator FSR without any mode hopping.

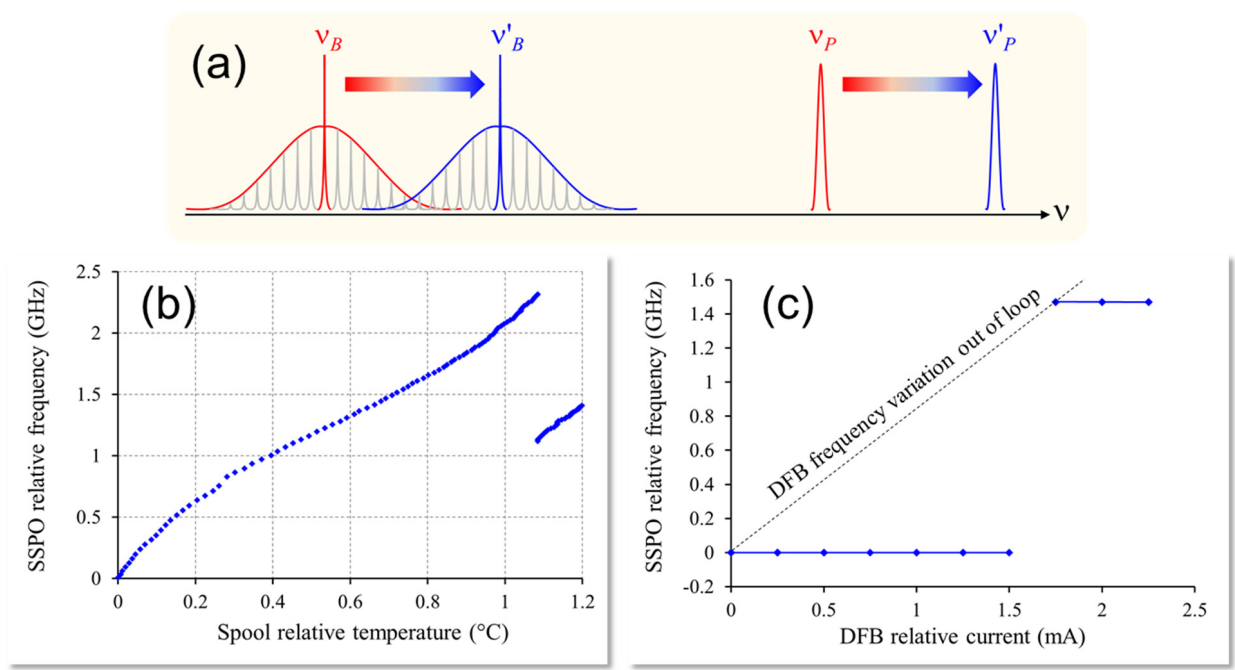

**Fig. 6: Evidence of self-adaptive oscillation.** **(a)** Frequency dragging mechanism. When the Stokes wave is frequency shifted, the pump wave follows this shift and consequently the same holds for the induced Brillouin gain. This mechanism prevents any mode hopping. In **(a)** and **(b)** the SAPO relative frequency is obtained from two SAPOs beating together where one SAPO only is perturbed. **(b)** Experimental evidence of the frequency dragging mechanism. On one SAPO, the fiber Brillouin resonator temperature is varied. The SAPO frequency is thus dragged over 500 FSRs without any mode hop. **(c)** When the DFB laser current is varied, the SAPO frequency remains constant although this perturbation is equivalent to 700 FSRs of the Brillouin resonator. These two features, shown in (b) and (c), shed light on the reason why a SAPO offers such robust single frequency oscillation.

By contrast, if a perturbation is now applied to the DFB current, its frequency remains unchanged since this frequency is locked by the Stokes wave frequency, i.e., the fiber resonator. This second mechanism is shown in Fig. 6(b) where the SAPO frequency stays constant while the DFB laser current is varied over 1.5 mA corresponding to 1.5 GHz of frequency variation of the out of loop DFB. Such perturbation is equivalent to 700 FSRs of the fiber resonator. These results show clearly that the overall stability of the SAPO is ruled by the stability of the long fiber Brillouin resonator. Moreover, they highlight that the observed 10-20 MHz frequency random walk is consistent with the 10 mK thermalization precision of the fiber spool.

## 6. Conclusion

A self-adaptive photonic oscillator based on the OEO paradigm is proposed. Its operation principle is presented and the key elements for its realization are emphasized. This system exhibits an extremely low frequency noise without any electronic servo loop. The Lorentzian linewidth is measured to be 40 mHz and the flicker noise linewidth is 200 Hz for 0.1 s integration time in free running condition. The long-term drift of the optical line is within 10 MHz over a few hours. Moreover, its intensity noise turned to be limited by the shot noise. Plug and play prototypes of this SAPO have been already developed in the laboratory. Their performances remain unchanged for operating temperatures ranging from -10°C to 40°C. Fine tunability can be achieved by stretching or thermo-controlling the Brillouin fiber resonator, whereas coarse tunability can be obtained by electrically or thermally controlling the DFB laser. Current developments address its locking to molecular references. Owing to its high spectral performances in a self-sustained manner to its operation robustness and to its implementation simplicity, SAPOs open a wide avenue to all the fields where narrow linewidth lasers are requested. All the more so SAPO principle can be easily extended to any wavelength between 700 nm and 2 μm where fibered components and semiconductor are currently available.